\documentclass{aa}
\defcitealias{K24}{KKV24}
\defcitealias{KF24}{KF24}
\usepackage{graphicx}
\usepackage{txfonts}
\usepackage[colorlinks=true,linkcolor=blue,citecolor=blue,filecolor=blue,urlcolor=blue]{hyperref}
\newcommand{\f}[1]{\fontfamily{qcr}\selectfont{#1}}
\begin{document}

   \title{Redistribution of ices between grain populations in protostellar envelopes}
   \subtitle{Only the coldest grains get ices}
      \author{
                Juris Kalv\=ans
                }
   \institute{
Engineering Research Institute ``Ventspils International Radio Astronomy Centre'' of Ventspils University of Applied Sciences,\\
In$\rm \check{z}$enieru 101, Ventspils, LV-3601, Latvia\\
\email{juris.kalvans@venta.lv}
                        }
   \date{Received October 8, 2024; accepted December 23, 2024}
  \abstract
{Matter that falls onto a protoplanetary disk (PPD) from a protostellar envelope is heated before it cools again. This induces sublimation and subsequent re-adsorption of ices that accumulated during the prestellar phase.}
{We explore the fate of ices on multiple-sized dust grains in a parcel of infalling matter.}
{A comprehensive kinetic chemical model using five grain-size bins with different temperatures was applied for an infalling parcel. The parcel was heated to 150\,K and then cooled over a total timescale of 20\,kyr. Effects on ice loss and re-accumulation by the changed gas density, the maximum temperature, the irradiation intensity, the size-dependent grain temperature trend, and the distribution of the ice mass among the grain-size bins were investigated.}
{A massive selective redistribution of ices exclusively onto the surface of the coldest grain-size bin occurs in all models. The redistribution starts already during the heating stage, where ices that are sublimated from warmer grains re-adsorb onto colder grains before complete sublimation. During the cooling stage, the sublimated molecules re-freeze again onto the coldest grains. In the case of full sublimation, this re-adsorption is delayed and occurs at lower temperatures because a bare grain surface has lower molecular desorption energies in our model.}
{Most protostellar envelope grains enter the PPD ice poor (bare). Ices are carried by a single coldest grain-size bin, here representing 12\,\% of the total grain surface area. This bare ice-grain dualism can affect the rate of the grain coagulation. The ice components are stratified on the grains according to their sublimation temperatures.}
\keywords{astrochemistry -- molecular processes -- protoplanetary disks -- planets and satellites: formation -- interstellar medium: dust -- stars: formation}
   \maketitle

\section{Introduction}
\label{intrd}

Planets around low-mass stars are formed from dust grains that consist of refractory materials and volatile ices. The grains initially reside in the interstellar medium, part of which transforms into a dense molecular cloud that contains a prestellar core \citep{Tafalla18}. After the protostar is born, the remainder of the core forms a protostellar envelope. Dust and gas fall from the envelope onto the protoplanetary disk (PPD). During this process, dust and gas are heated, up to a few hundred K. This can occur via two primary mechanisms. Firstly, heating is provided by a mild accretion shock at the centrifugal barrier near the outer rim of the PPD \citep{Cassen81,Sakai14}. Secondly, when the infall trajectory is near the outflow cavity, the grains are temporarily heated by protostellar radiation \citep{Visser09}. Upon arrival on the disk, the gas and dust become shielded from protostellar and interstellar radiation and cool again. The disk material continues to experience heating events either through shocks in the disk or through flaring of the protostar \citep{Armitage11}.

The composition of interstellar ices can be reset during star formation before the ices are included into planets or comets \citep{Jorgensen20}. The partial resetting of ice before it arrives in PPDs has been confirmed by astrochemical simulations \citep{Visser11,Drozdovskaya14,Aota15,Furuya17} and in PPDs themselves \citep{Podolak11,Molyarova18}. These studies employed models that considered a single grain size. However, grains are expected to have a variety of sizes, shapes, and composition. Their size and material can cause differences in the dust grain temperatures $T_d$ . A proven approach in numerical simulations is to discern the grains according to their size \citep{Acharyya11}. For the modelling, the grain-size distribution can be represented by several discrete grain-size bins. These multi-grain models show that different temperatures for the grain-size bins affect the formation rate of the icy mantles and their eventual composition \citep{Pauly16}.

The temperatures experienced by a gas-dust parcel infalling from the envelope towards the PPD can exceed the sublimation temperatures of the main ice species CO, CO$_2$, and H$_2$O. In this case, ice species first sublimate from the hottest grain population. \citet[][hereafter KF24]{KF24} found that observations of complex organic molecules (hot corinos) and carbon chains (warm carbon-chain chemistry) in low-mass protostars are best explained by an assortment of temperatures of the dust grains, rather than a single $T_d$ for all grains.

The aim of this study is to explore how the rise and decrease in grain temperatures at typical infall conditions redistribute the ices between hotter and colder grain populations. In particular, we explore whether a heating-cooling cycle can induce an accumulation of ices in the coldest grain-size bin but leaves other grains ice poor. This situation briefly occurred during heating in the \citetalias{KF24} model. An asymmetric redistribution of ices like this can considerably affect the grain size and stickiness, which in turn affects the expected starting conditions for planet formation in PPDs. To fulfil the aim of this study, we applied the multi-grain model of \citetalias{KF24} for a parcel of circumstellar matter infalling towards the inner region of a protostellar core. The parcel was heated before it cooled again, presumably during infall towards the inner region (PPD) of a protostellar core.

\section{Methods}
\label{mthd}

The study is based on the model of \citet[][hereinafter KKV24]{K24}. The application of this model for protostellar envelope chemistry has illustrated the importance of the grain temperature distribution in circumstellar environments \citepalias{KF24}.

\subsection{Chemical model}
\label{micr}

\begin{table}
\caption{Features and parameters of the {\f Standard} model.}
\label{tab-std}
\small
\begin{tabular}{l l l l}
\hline\hline
Name & Parameter & Value & Notes, refs. \\
\hline
Gas density & $n_H$\,[cm$^{-3}$] & $10^8$ &  \\
Interstellar extinction & $A_V$\,[mag] & 10 &  \\
Gas temperature & $T_{\rm gas}$\,[K] & ... &  \\
Dust temperature & $T_d$\,[K] & ... &  \\
$T_d$ exponent & $y$ & -1/6 & Eq.~\ref{micr1} \\
Initial $T_{\rm gas}$ & $T_{\rm ini}$\,[K] & 10 & \\
Maximum $T_{\rm gas}$ & $T_{\rm max}$\,[K] & 150 &  \\
Final $T_{\rm gas}$ & $T_{\rm end}$\,[K] & 40 &  \\
Heating time to $T_{\rm max}$ & $t_{\rm heat}$\,[yr] & $10^4$ &  \\
Time at $T_{\rm max}$ & $t_{\rm plat}$\,[yr] & 0 &  \\
Cooling time to $T_{\rm end}$ & $t_{\rm cool}$\,[yr] & $10^4$ &  \\
CR ionization rate & $\zeta_{\rm CR}\,[s^{-1}]$ & $4\times10^{-17}$ & 1, 2 \\
Irradiation factor & $I$ & 1.0 &  \\
Gas network &  & \textsc{rate12} & 3\tablefootmark{a} \\
Surface network &  & OSU & 4\tablefootmark{b} \\
Initial abundances &  & M-1/6COM & 5\tablefootmark{c}, Table\,\ref{tab-ab} \\
Grains/gas &  & 0.5\,\% & by mass \\
Ice/gas &  & 0.8\,\% & by mass\tablefootmark{d} \\
Grain size bins &  & 5 & 2 \\
\hline
\end{tabular}
\tablefoottext{a}{http://udfa.ajmarkwick.net}
\tablefoottext{b}{Reduced to match \textsc{rate12}.}
\tablefoottext{c}{At the end of the prestellar stage.}
\tablefoottext{d}{At total freeze-out.}
\tablebib{(1) \citet{Ivlev15}, (2) \citetalias{K24}, (3) \citet{McElroy13}, (4) \citet{Garrod08}, (5) \citetalias{KF24}.}
\end{table}
%

\begin{table}
\caption{Basic data of the five grain populations.}
\label{tab-gr}
\small
\begin{tabular}{c c c c c}
\hline\hline
$a$ [$\rm \mu$m] & $n_d/n_H$\tablefootmark{a} & $b$ [MLs] & $T_{d,\rm ini}$\tablefootmark{b}\,[K] & $T_{d,\rm max}$\tablefootmark{c}\,[K] \\
\hline
0.037 & $5.46\times10^{-12}$ & 82 & 10.8 & 177 \\
0.058 & $1.73\times10^{-12}$ & 76 & 10.3 & 164 \\
0.092 & $5.46\times10^{-13}$ & 72 & 9.8 & 152 \\
0.146 & $1.73\times10^{-13}$ & 68 & 9.2 & 141 \\
0.232 & $5.46\times10^{-14}$ & 64 & 8.6 & 125 \\
\hline
\end{tabular}
\tablefoottext{a}{Number density relative to H atoms.}
\tablefoottext{b}{Initial $T_d$ at $T_{\rm gas}=10$\,K.}
\tablefoottext{c}{Maximum $T_d$ at $T_{\rm gas}=150$\,K.}
\end{table}
%

\begin{table}
\caption{Elemental abundances relative to hydrogen.}
\label{tab-elem}
\centering
\begin{tabular}{c l}
\hline\hline
Species & X/H \\
\hline
H & 1.00\\
He & 0.090 \\
C & $1.4\times10^{-4}$ \\
N & $7.5\times10^{-5}$ \\
O & $3.2\times10^{-4}$ \\
F & $6.7\times10^{-9}$ \\
Na & $2.0\times10^{-9}$ \\
Mg & $7.0\times10^{-9}$ \\
Si & $8.0\times10^{-9}$ \\
P & $3.0\times10^{-9}$ \\
S & $8.0\times10^{-8}$ \\
Cl & $4.0\times10^{-9}$ \\
Fe & $3.0\times10^{-9}$ \\
\hline
\end{tabular}
\end{table}
%

\begin{table}
\caption{Initial X/H of some major species.}
\label{tab-ab}
\tiny
\begin{tabular}{l c c c c c c}
\hline\hline
 & \multicolumn{6}{l}{X/H per grain size bin [$\rm\mu$m]} \\
Species & 0.037 & 0.058 & 0.092 & 0.146) & 0.232 & Total \\
\hline
H$_2$(gas) & ... & ... & ... & ... & ... & 0.500 \\
H(gas) & ... & ... & ... & ... & ... & 4.7(-9)\tablefootmark{a} \\
H$_2$O(ice) & 5.4(-5) & 3.5(-5) & 2.5(-5) & 1.9(-5) & 1.4(-5) & 1.5(-4) \\
CO(ice) & 5.1(-5) & 2.3(-5) & 1.2(-5) & 6.2(-6) & 3.3(-6) & 9.5(-5) \\
N$_2$(ice) & 1.5(-5) & 8.0(-6) & 4.8(-6) & 3.1(-6) & 2.0(-6) & 3.3(-5) \\
CO$_2$(ice) & 8.2(-6) & 5.5(-6) & 4.1(-6) & 3.1(-6) & 2.4(-6) & 2.3(-5) \\
CH$_3$OH(ice) & 5.3(-6) & 3.0(-6) & 2.0(-6) & 1.4(-6) & 1.1(-6) & 1.3(-5) \\
NH$_3$(ice) & 3.2(-6) & 2.1(-6) & 1.5(-6) & 1.1(-6) & 8.6(-7 & 8.8(-6) \\
\hline
\end{tabular}
\\
\tablefoottext{a}{A(B) means $\rm A\times10^{\rm B}$.}
\end{table}

The recently updated multi-grain kinetic rate-equation astrochemical model \textsc{Alchemic-Venta} \citep[originally introduced by][]{K15} now has an advanced treatment for molecular grain surface processes. We only include a brief description of the chemical aspects and refer to \citetalias{K24}, where the chemical aspects are explained in detail and with the relevant references. Table~\ref{tab-std} summarises the features of the code and the default values of the important physical and chemical parameters, as used in our {\f Standard} model.

A single major change was introduced in the code: The bulk-ice multi-layer approach for modelling the icy mantles on grains was removed. This was required for a physically correct description of rapid thermal desorption and re-adsorption of ice species onto grains. With a multi-layer approach, the rate of these processes could be artificially slowed down by the simulated transfer of molecules between the ice layers. Considering bulk ice does not substantially change the overall surface chemistry at elevated temperatures of the protostellar stage \citep{Garrod13}.

The surface chemistry starts with molecule adsorption onto the grain surface. Ice accumulates into five grain-size bins, and smaller grains have larger populations according to the MRN grain-size distribution, as shown in Table~\ref{tab-gr}. All grains share a similar composite chemical make-up. The grains were assumed to be partially processed, as expected for a star-forming region. More precisely, the small grains were depleted onto the larger grains \citep{Silsbee20,Sipila20}, whose abundance was increased by a factor of 1.25 to retain the MRN mass budget. Moreover, the surfaces of the refractory grains were assumed to be carbonaceous to account for the adsorption of polycyclic aromatic hydrocarbon molecules and polymerised organic residue that formed by the over-photoprocessing of ices during the previous star-forming activity \citep{Hagen79,Allamandola88,Quirico16}. This assumed carbonaceous coating is not able to participate in strong inter-molecular hydrogen bonds with ice molecules, and this affects the chemical properties of bare grains \citepalias{K24}.

The average temperature of dust depends on its composition, on the intensity of electromagnetic radiation, and on the grain size \citep{Li01,Draine03,Cuppen06,Li12}. The latter two aspects were considered here. The temperature of the grain populations was determined by the grain radius, following the equation
\begin{equation}
        \label{micr1}
T_{d,a} = \left(\frac{a+b}{0.1}\right)^y \times T_{d,0.1}\,,
\end{equation}
where $a$ is the radius of the refractory grain, $b$ is the ice-mantle thickness (in $\rm\mu$m), and $T_{d,0.1}$ is the temperature of 0.1\,$\rm\mu$m grains \citep{Krugel07,Pauly16}. The exponent $y$ is a variable. For simplicity, $T_{d,0.1}$ was assumed to be equal to $T_{\rm gas}$ (Sect.~\ref{grd}). $T_{\rm gas}$ has been shown to be similar to $T_d$ of sub-micron grains by more complex models \citep{Gavino21}.

The surface processes include binary reactions, photodissociation of the ice species, and six desorption mechanisms. These are sublimation, photodesorption by the interstellar UV radiation field (ISRF), and cosmic-ray (CR) induced photons, chemical desorption of the products of exothermic reactions, indirect chemical desorption by the frequent H+H$\rightarrow$H$_2$ surface reaction, and CR-induced whole-grain heating. Out of these, sublimation primarily matters in the protostellar stage, while the other processes are mainly important for creating a feasible composition of the ice in the prestellar core that is inherited by the protostellar envelope. The rate of binary reactions and desorption is affected by the molecular desorption energy $E_D$, which varies depending on the surroundings of the molecule, whether it is a bare grain, strongly polar (H$_2$O), or hyper-volatile ices (CO).

As a dedicated chemical model, \textsc{Alchemic-Venta} has benefits for the calculation of the grain-size (ice-mantle thickness) evolution. It includes the UMIST \textsc{rate12} reaction network\footnote{http://udfa.ajmarkwick.net} and considers ice photoprocessing and desorption by several mechanisms, with rate coefficients that are updated according to recent experimental and theoretical results \citepalias{K24}. The most significant chemical aspect probably is the solid-phase conversion of CO into CO$_2$ and CH$_3$OH, that is, the chemical production of species with increasing sublimation temperatures $T_{\rm subl}$, which is not easily reproduced by simplified functions \citep[e.g.,][]{Rawlings21}. This conversion is important because when the gas temperature $T_{\rm gas}$ rises, it allows repeated freeze-out of a major part of the C and O atoms into ice after the sublimation of the CO ice. Other chemical transformations of similar nature include $\rm N_2\rightarrow NH_3$ and the $\rm CH_4\rightarrow C_2H_6\rightarrow$ longer carbon chains \citepalias{KF24}.

We studied events in a protostellar envelope that inherited its ices from the dark prestellar core. Table~\ref{tab-elem} lists the elemental abundances. Therefore, the initial conditions were set to be characteristic for these cores, with $T_{\rm gas}\equiv T_{d,0.1}=10$\,K, and the initial chemical composition was drawn from the prestellar stage model of \citetalias{KF24}, using the elements of Table~\ref{tab-elem}. Table~\ref{tab-ab} lists the initial abundance of the most important molecules (we used abundances relative to H (X/H), as used in the models, while in the results sections, the abundances are relative to H$_2$, $n/n(\rm H_2$), as is often shown for astrochemical observations). The ISRF extinction was kept constant at $A_V=10$\,mag. The effects of the ISRF photons are still insignificant for the {\f Standard} model. The number density of H atoms was also kept constant, with $n_H=10^8$\,cm$^{-3}$ as the default value.

According to Eq.\,(\ref{micr1}), $T_d$ smoothly grows for all grains with increasing $T_{d,0.1}$, while the difference between $T_d$ for grains in various size bins increases consistently. This may not always be the case as the incident spectrum of protostellar radiation is modified by absorption and scattering. Variations in the spectrum and intensity due to distance to the protostar and the amount of intervening material may change the relative efficiency for the heating of grains of different sizes. Furthermore, the optical properties of the grains also change when ices sublimate or re-freeze onto their surfaces. These effects were not accounted for in our model and can affect the relative temperatures for grains of different sizes.

\subsection{Model parameters}
\label{grd}

\begin{table}
\caption{Investigation models and their changed parameters relative to the {\f Standard} model.}
\label{tab-grd}
\small
\begin{tabular}{l l l l l l l}
\hline\hline
 & Changed &  & {\f Standard} \\
Model & parameters & Value & Value \\
\hline
{\f Dens6} & $n_H$\,[cm$^{-3}$] & $10^6$ & $10^8$ \\
{\f Dens7} & $n_H$\,[cm$^{-3}$] & $10^7$ & $10^8$ \\
{\f Dens10} & $n_H$\,[cm$^{-3}$] \& $y$ & $10^{10}$ \& -1/20 & $10^8$ \& -1/6 \\
{\f Dens12} & $n_H$\,[cm$^{-3}$] \& $y$ & $10^{12}$ \& -1/20 & $10^8$ \& -1/6 \\
{\f Tmax50} & $T_{\rm max}$ \& $T_{\rm end}$\,[K] & 50 \& 20 & 150 \& 40 \\
{\f Tmax100} & $T_{\rm max}$ \& $T_{\rm end}$\,[K] & 100 \& 30 & 150 \& 40 \\
{\f Plateau} & $t_{\rm plat}$\,[kyr] & 10 & 0 \\
{\f Irrad} & $I$ & 100 & 1 \\
{\f Tpow1/6} & $y$ & 1/6 & -1/6 \\
{\f Tpow-1/20} & $y$ & -1/20 & -1/6 \\
{\f Ice\_small} & \multicolumn{2}{l}{(all ice initially on 0.037\,$\rm\mu$m grains)} & \\
{\f Ice\_large} & \multicolumn{2}{l}{(all ice initially on 0.232\,$\rm\mu$m grains)} & \\
\hline
\end{tabular}
\end{table}

\citet{Visser09,Drozdovskaya14}, and \citet{Furuya17} indicated that for an infalling parcel of matter in a low-mass protostellar envelope, the rise and fall of the temperature occurs on timescales of about 20--100\,kyr. We adopted the shorter timescale, divided equally between the heating ($t_{\rm heat}$) and cooling ($t_{\rm cool}$) episodes, as listed in Table~\ref{tab-std}. This is a conservative assessment because longer timescales would favour sublimation of ices and thus a more complete redistribution to colder grain-size bins. Therefore, $t_{\rm heat}$ and $t_{\rm cool}$ were not considered as variables in this study. The temperature increased from $T_{\rm ini}=10$\,K to the peak gas temperature $T_{\rm max}$ over 10\,kyr proportionally to the integration time squared $t^2$ \citep{Garrod08}, while the decrease from $T_{\rm max}$ to the final temperature of $T_{\rm end}=40$\,K followed a reverse curve. Because $T_{\rm gas}$ did not return to 10\,K, the total integration time for the up-down temperature curve was 16.2\,kyr, not 20\,kyr. To study what happens after the heating episode, the simulations were continued for another 4\,kyr with $T_{\rm gas}=T_{\rm end}$, up to a total time-span of 20\,kyr.

We assumed that the parcel did not enter dense, cold, and shielded inner parts of the PPD during the simulation. Thus, $A_V$ and $n_H$ did not rise significantly, and $T_{\rm end}$ remained relatively high at 40\,K in Model {\f Standard}. Possible mixing with other infalling matter (nearby parcels) was not considered. The simplified conditions with fixed $A_V$, $n_H$, $T_{\rm max}$, $t_{\rm heat}$, and $t_{\rm cool}$ were not intended to reflect conditions in real infalling protostellar gas. Instead, the fixed conditions, roughly representative of protostellar envelopes, and the simple $T_d$ dependence on the grain size allowed us to understand the transfer of volatile matter through gas between grain populations with different $T_d$ without other complex background physics. In addition to the {\f Standard} model, described above and in Table~\ref{tab-std}, a number of other simulations were carried out. A physical parameter, $n_H$, $T_{\rm max}$ (and $T_{\rm end}$), the irradiation intensity factor $I$, the time spent at an elevated temperature $t_{\rm plat}$, $y$, or the initial abundance of ices, was modified in each of these models relative to {\f Standard}. Table~\ref{tab-grd} lists all the models and the changes in their parameter values. The considerations for the selection of changed parameters and their values are described below in Sect.~\ref{rslt}, along with the modelling results.

\section{Results}
\label{rslt}

\begin{table*}
\caption{Gas temperatures for an ice loss of 50\,\% relative to the initial abundances ($T_{\rm des}$\,[K]) during the heating stage and a re-adsorption of 50\,\% ($T_{\rm re}$\,[K]) during the cooling stage for six major ice species.}
\label{tab-50}
\centering
\begin{tabular}{l |  l l |  l l |  l l |  l l |  l l |  l l }
\hline\hline
 & \multicolumn{2}{c}{N$_2$} & \multicolumn{2}{c}{CO} & \multicolumn{2}{c}{CO$_2$} & \multicolumn{2}{c}{CH$_3$OH} & \multicolumn{2}{c}{NH$_3$} & \multicolumn{2}{c}{H$_2$O} \\
Initial $n/n\rm(H_2)$ & \multicolumn{2}{c}{$6.5\times10^{-5}$} & \multicolumn{2}{c}{$1.9\times10^{-4}$} & \multicolumn{2}{c}{$4.6\times10^{-5}$} & \multicolumn{2}{c}{$2.5\times10^{-5}$} & \multicolumn{2}{c}{$1.8\times10^{-5}$} & \multicolumn{2}{c}{$2.9\times10^{-4}$} \\
Model & $T_{\rm des}$ & $T_{\rm re}$ & $T_{\rm des}$ & $T_{\rm re}$ & $T_{\rm des}$ & $T_{\rm re}$ & $T_{\rm des}$ & $T_{\rm re}$ & $T_{\rm des}$ & $T_{\rm re}$ & $T_{\rm des}$ & $T_{\rm re}$ \\
\hline
{\f Dens6} & 21 & … & 24 & … & 54 & 40 & 104 & … & 113 & 50 & 116 & 51 \\
{\f Dens7} & 22 & … & 26 & … & 60 & 61 & 127 & 73 & 143 & 78 & 145 & 78 \\
\textbf{{\f Standard}} & \textbf{26} & \textbf{…} & \textbf{31} & \textbf{…} & \textbf{70} & \textbf{72} & \textbf{139} & \textbf{138} & \textbf{150} & \textbf{148} & \textbf{…} & \textbf{…} \\
{\f Dens10} & 26 & … & 30 & … & 72 & 70 & 135 & 92 & 145 & 92 & 147 & 92 \\
{\f Dens12} & 30 & … & 34 & … & 76 & 80 & … & … & … & … & … & … \\
{\f Tmax50} & 26 & 26 & 31 & 29 & … & … & … & … & … & … & … & … \\
{\f Tmax100} & 26 & … & 31 & 30 & 67 & 66 & … & … & … & … & … & … \\
{\f Plateau} & 26 & … & 31 & … & 70 & 51 & 139 & … & 150 & 96 & 150 & 96 \\
{\f Irrad100} & 26 & … & 18 & … & 69 & 60 & 122 & … & … & … & 150 & 146 \\
{\f Tpow1/6} & 24 & … & 28 & … & 64 & 63 & 123 & 95 & 134 & 95 & 138 & 95 \\
{\f Tpow-1/20} & 23 & … & 27 & … & 60 & 63 & 117 & 83 & 131 & 84 & 132 & 84 \\
{\f Ice\_small} & 26 & … & 31 & … & 66 & 72 & 107 & 94 & 114 & 94 & 114 & 95 \\
{\f Ice\_large} & 28 & … & 33 & … & 73 & 72 & 141 & 138 & 150 & 147 & … & … \\
\hline
\end{tabular}
\end{table*}

The gas-grain chemistry of the prestellar stage produces icy mantles that are inherited by the protostellar envelope. To study the sublimation of ices, it is convenient to classify the ice species into groups that share similar sublimation temperatures. These groups of species are primarily the hyper-volatile species N$_2$, CO, and CH$_4$ ices, which already sublimate at $T_d>22$\,K, CO$_2$ ice that sublimates at $T_d>55$\,K, and NH$_3$, CH$_3$OH, and H$_2$O ices, which start to sublimate when $T_d$ exceeds about 90\,K. The latter two groups include a variety of other compounds, such as C$_2$H$_2$, H$_2$S, and HCN, which sublimate together with CO$_2$, as well as HCOOH, H$_2$O$_2$, and HF, which sublimate together with or after H$_2$O ice. In what follows, the primary focus is on the total abundance of the species that constitute the groups and their chemical transformation into molecules that belong to another group.

The modelling of hundreds of species and thousands of reactions allowed us to simulate a series of chemical changes in the species and their phase transition that affect the grain-size evolution, such as the partial re-freeze of sublimated CO in the form of CO$_2$, the re-freeze of N$_2$ in the form of NH$_3$, and the re-freeze of CO$_2$ oxygen in the form of H$_2$O. At the end of the prestellar stage, all grain sizes had a $b$ of about 64...82 ice molecule monolayers (MLs; Table~\ref{tab-gr}). A few percent of C and O are in many organic species that contain several heavy atoms, each with abundances relative to H$_2$ ($n/n\rm(H_2)$) below 10$^{-6}$.

\subsection{Standard results}
\label{r-std}

\begin{figure}
\includegraphics{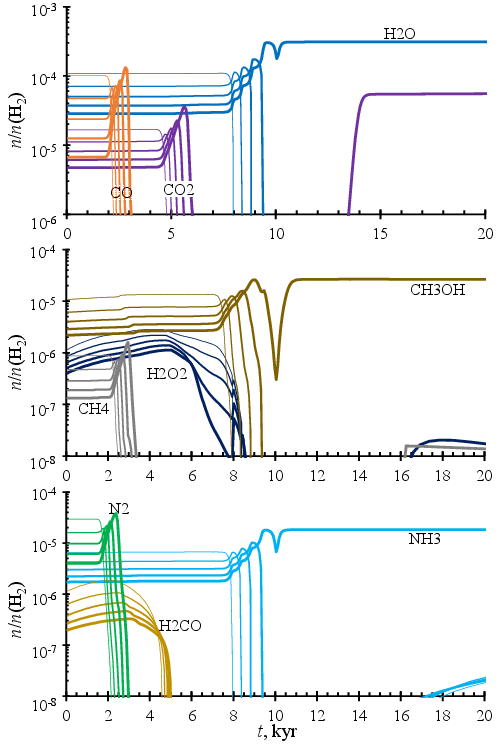}
        \vspace{-17.5cm}
        \caption{Evolution of $n/n\rm(H_2)$ for the major ice molecules on the surfaces of the five grain-size bins in the {\f Standard} model. The thicker lines show larger grains.}
        \label{fig-grab}
\end{figure}

We start with the results of the {\f Standard} model. These were further used as a reference for the comparison with other simulations. The ice sublimation started with the desorption of dinitrogen N$_2$. Half or 50\,\% of the N$_2$ ice relative to its initial abundance was thermally desorbed at an integration time $t=2.4$\,kyr and a gas temperature $T_{\rm des}=26$\,K. The sublimation was not momentary; 99\,\% N$_2$ ice disappeared in the $T_{\rm gas}$ interval between 19 and 29\,K ($t=1.5...2.7$\,kyr), while 80\,\% N$_2$ ice loss was observed at 23...28\,K (2.1...2.7\,kyr). All ice species showed sublimation periods of several $10^2$\,yr, with a temperature rise to about 10\,K. These slow sublimation processes occur because the grain temperatures are spread out and we did not use a single $T_d$ value.

The 40\,K value for $T_{\rm end}$ in the {\f Standard} model is too high for N$_2$ to re-freeze again; thus it does not have a value for its 50\,\% re-adsorption temperature $T_{\rm re}$. The ice molecule that experienced re-accumulation during the cooling stage in almost each simulation was carbon dioxide, CO$_2$. In the {\f Standard} model, the temperature $T_{\rm re}$ for 50\,\% CO$_2$ re-adsorption (relative to its initial abundance) is $T_{\rm re,CO2}=72$\,K at $t=13.9$\,kyr. Re-accumulation of CO$_2$ ice from 10\,\% to 90\,\% of its initial abundance occured from 75 to 69\,K (13.7...14.1\,kyr), while freeze-out from 1\,\% to 99\,\% of the initial abundance occured in the 80...68\,K interval (13.4...14.2\,kyr). These numbers show that ice re-accumulation also takes several hundred years. In the case of CO$_2$, the ice accumulation continued and reaches 120\,\% relative to its initial abundance because it is produced via surface reactions. Similarly to N$_2$, carbon monoxide, CO, remained in the gas-phase at the end of model {\f Standard}. The abundance of CO slowly decreased as it was converted into CO$_2$ ice. This change is insignificant in model {\f Standard} (only 2\,\% of CO), but becomes a major CO sink under irradiated conditions (Sect.~\ref{r-irr}). Table~\ref{tab-50} lists the $T_{\rm des}$ and $T_{\rm re}$ values for all models, along with the initial $n/n\rm(H_2)$ for all the major ice species N$_2$, CO, CO$_2$, methanol, CH$_3$OH, ammonia, NH$_3$, and water ice, H$_2$O. As discussed below, $T_{\rm des}$ and $T_{\rm re}$ are key parameters in understanding what occurred in each simulation.

Figure~\ref{fig-grab} shows that the sublimation of ices occurs sequentially. It starts in the hottest grain-size bin. After they are sublimated from a hot grain-size bin, the former icy molecules quickly re-freeze onto grains whose $T_d$ has not yet reached their thermal desorption temperature. Thus, the coolest grains accumulate most of the molecules that were sublimated from other grain-size bins, until the coolest grains are too hot and also sublimate their ices. This process is repeated for the hyper-volatile species CO, N$_2$, and CH$_4$ and for less volatile species, such as CO$_2$. In the case of H$_2$O, NH$_3$, and CH$_3$OH, this process occurs near $T_{\rm max}$, and the sublimation from the largest, coldest grains is interrupted by the start of the cooling stage. Reactive species, such as hydrogen peroxide, H$_2$O$_2$, and formaldehyde, H$_2$CO, are initially produced via photoprocessing (of H$_2$O and CH$_3$OH ices, respectively) and are mostly destroyed at elevated temperatures even before full sublimation.

Figure~\ref{fig-dens} shows that the sequential sublimation from the five grain-size bins produces a characteristic seesaw pattern for the gas-phase abundances of the less volatile molecules CH$_3$OH, NH$_3$, and H$_2$O in the heating stage of the {\f Standard} model \citepalias{KF24}. Table~\ref{tab-gr} lists the temperatures of the five grain-size bins when $T_{\rm gas}=T_{\rm max}$. As the temperature drops, only H$_2$O, NH$_3$, CH$_3$OH, and CO$_2$ re-freeze before $T_{\rm end}=40$\,K. The re-formation of icy mantles exclusively occurs onto the largest, coldest grains. For these, the ice thickness $b_{0.232}$ reaches 363\,MLs at the end of the simulation ($t=20$\,kyr), while it is below 1\,ML for all the other grain-size bins.

The abundances of the re-accumulated ices are slightly changed compared to their initial values, as listed in Table~\ref{tab-50}: 106\,\% for H$_2$O, 102\,\% for CH$_3$OH, 105\,\% for NH$_3$, and 120\,\% for CO$_2$ ice. The increase arises from CO and N$_2$, which are converted into less volatile molecules that can be stored on icy grains. Similar solid-phase final abundances were also obtained for other models, unless noted otherwise.

\subsection{Gas density}
\label{r-dens}

\begin{figure*}
\includegraphics{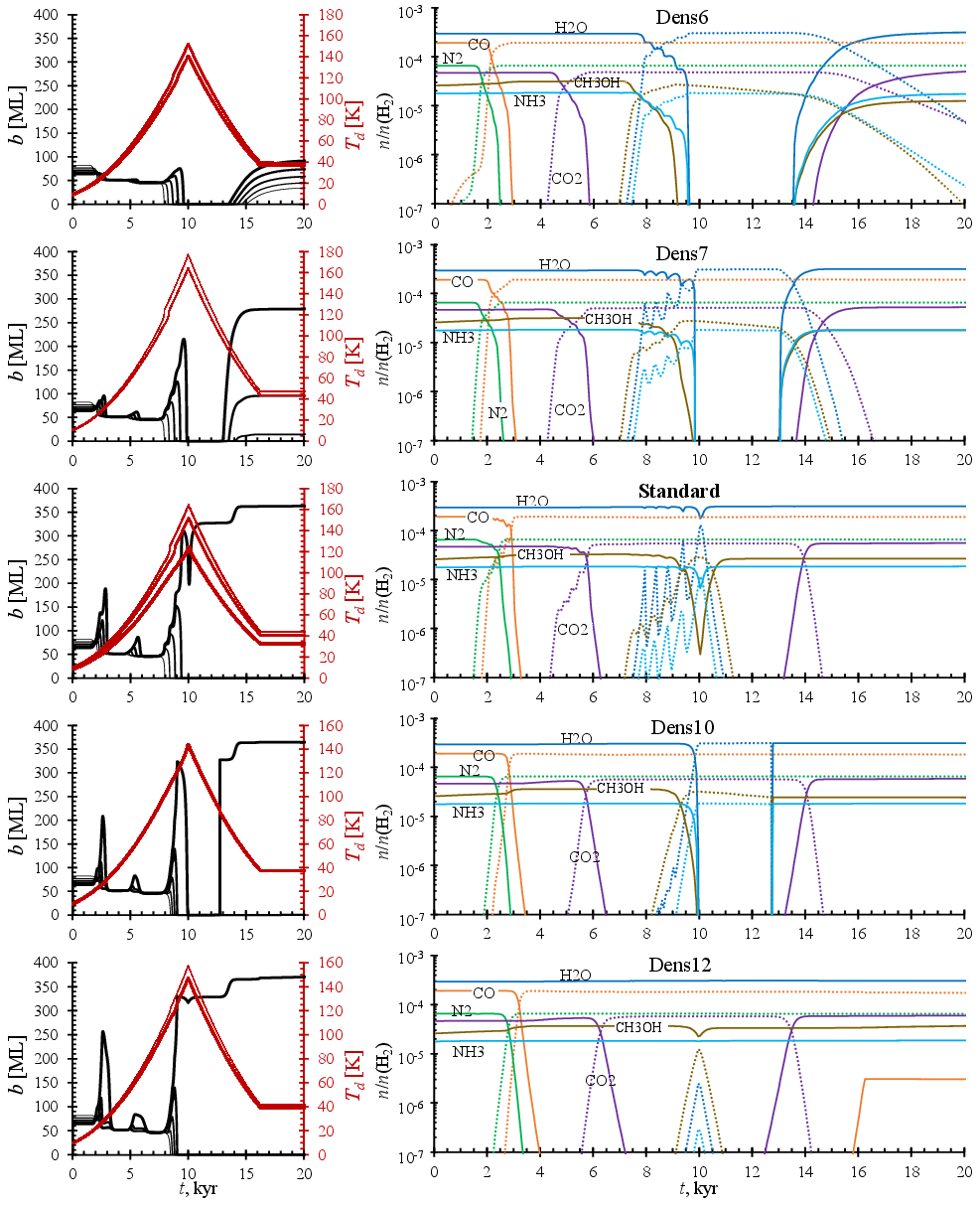}
        \vspace{-9.7cm}
        \caption{Effects of $n_H$ on the ice sublimation and re-freeze with Models {\f Dens6}, {\f Dens7}, {\f Standard}, {\f Dens10}, and {\f Dens12} from top to bottom. Left: Ice thickness (black lines) changes with $T_d$ (red lines in the background). The thicker lines show larger grains. Right: $n/n\rm(H_2)$ for ice (solid lines) and gas-phase (dotted lines) species.}
        \label{fig-dens}
\end{figure*}

The hydrogen atom number density $n_H$ has an assumed default value of $10^8$\,cm$^{-3}$ in Model {\f Standard}. Practically all of the H atoms are in H$_2$ molecules. The value of $n_H$ can reach about $10^{12}$\,cm$^{-3}$ during the transfer from the envelope to the disk \citep{Visser11}. To investigate how these higher densities affect the transfer of ices between grain-size bins, we also considered $n_H=10^{10}$\,cm$^{-3}$ in Model {\f Dens10} and $10^{12}$\,cm$^{-3}$ in model {\f Dens12}. In addition, observations of complex organic molecules and warm carbon-chains in protostellar envelopes \citep{Cazaux03,Sakai08} indicate that significant heating also occurs in less dense regions. For these, we used Models {\f Dens6} and {\f Dens7} with $n_H$ of $10^6$ and $10^7$\,cm$^{-3}$, respectively. 

In the higher-density Models {\f Dens10} and {\f Dens12}, which are relevant for PPD conditions, the $T_d$ differences for sub-$\rm\mu$m grains are expected to be smaller than those derived with $y=-1/6$ \citep{Podolak11}. To reflect a strong gas-grain thermal coupling, we assumed that $y=-1/20$ in these models. With this approach, the $T_d$ difference for the smallest and largest grains at $T_{\rm gas}=10$\,K is below 0.5\,K \citep[lower than that calculated with the sophisticated model of][]{Gavino21} and 16\,K at $T_{\rm gas}=150$\,K. The smaller differences in $T_d$ may enable re-adsorption of gas-phase volatiles onto the surfaces of other grain-size bins, not just the coldest-grain population, because the other grain-size bins could also have sufficiently low $T_d$.

With a higher $n_H$, the molecules more often stick to the grains, which allows a faster growth of the icy mantles. In this way, higher $n_H$ shifts the ice sublimation to higher temperatures. With $y$ assumed to be -1/20, the lowest $T_d$ is now higher relative to $T_{\rm gas}$, which induces sublimation at lower $T_{\rm gas}$. Table~\ref{tab-50} shows that volatiles sublimate at higher $T_{\rm gas}$ in Model {\f Dens10}, where even H$_2$O ice is sublimated, unlike in the {\f Standard} model. In Model {\f Dens12} the rate of molecule adsorption onto grains, which is proportional to ${n_H}^2$, is so high that no significant degradation of the H$_2$O ice abundance is observed. In other words, at a higher $n_H$, a higher $T_d$ is needed for the thermal desorption rate to be higher than the rate of molecules that stick to grains. Moreover, some re-freeze near the end of the simulation at $T_{\rm end}=40$\,K can also be experienced by the hyper-volatile molecules. This is shown by the re-accumulation of CO in Model {\f Dens12}, where the end abundance of CO ice reaches 2\,\% of its initial abundance. In the two high-density models, a full migration of the ices to the coldest, largest 0.232\,$\rm\mu$m grains occurs, and $b_{0.232}$ reaches 365 and 370\,MLs for {\f Dens10} and {\f Dens12}, respectively.

In Models {\f Dens6} and {\f Dens7}, the number density of gas and dust is lower by factors of 100 and 10 relative to the {\f Standard} model, respectively. The rate of molecules and atoms that stick to the grains is correspondingly lower by factors of 10,000 and 100. As a result, Fig.~\ref{fig-dens} shows that the coldest-grain population is unable to adsorb the volatiles before other grain-size bins become cold enough to grow their own icy mantles. In Model {\f Dens7} most of the ice mass is still located on the large $a=0.232\,\rm\mu$m grains with an ice thickness $b_{0.232}=279$\,MLs. The next largest bin with $a=0.146$\,$\rm\mu$m has $b_{0.146}=96$\,MLs, the 0.092\,$\rm\mu$m grains have $b_{0.092}=14$\,MLs, while for the two smallest and hottest grain-size bins (0.058 and 0.037\,$\rm\mu$m) $b$ is below 1\,ML. For Model {\f Dens6}, the ice mass is evenly distributed between all grain-size bins with an ice thickness between $b_{0.232}=91$\,MLs and $b_{0.037}=34$\,MLs. Table~\ref{tab-50} shows that the sublimation of ices and especially their re-accumulation occurs at lower temperatures for the low-density models.

Table~\ref{tab-50} shows that there is a major difference for $T_{\rm re}$ between the {\f Standard} model on one hand and Models {\f Dens6}, {\f Dens7}, and {\f Dens10} on the other hand. The reason for the low $T_{\rm re}$ values is that these models desorb all their ice at $T_{\rm max}$, exposing the bare-grain surface. The bare surfaces were assumed to be carbonaceous in our model, and did not form hydrogen bonds \citepalias{K24}. This means that the H$_2$O, NH$_3$, and CH$_3$OH surface molecules sublimate above $T_d\approx80$\,K. A significant adsorption of these molecules is prevented until the $T_d$ of the coldest grain population allows accumulating $\approx$1\,ML of ice. After this point, the ice accumulation is quick. An extreme case here is CH$_3$OH, which only re-freezes to 48\,\% of the initial abundance because the gas in Model {\f Dens6} is destroyed. On the other hand, most CH$_3$OH ice does not sublimate in Model {\f Dens12} and can therefore reach an abundance of 144\,\% relative to the initial value. The surplus abundance is reached during the first few thousand years because CO is hydrogenated on cold grains.

\subsection{Maximum temperature and its duration}
\label{r-temp}

\begin{figure*}
\includegraphics{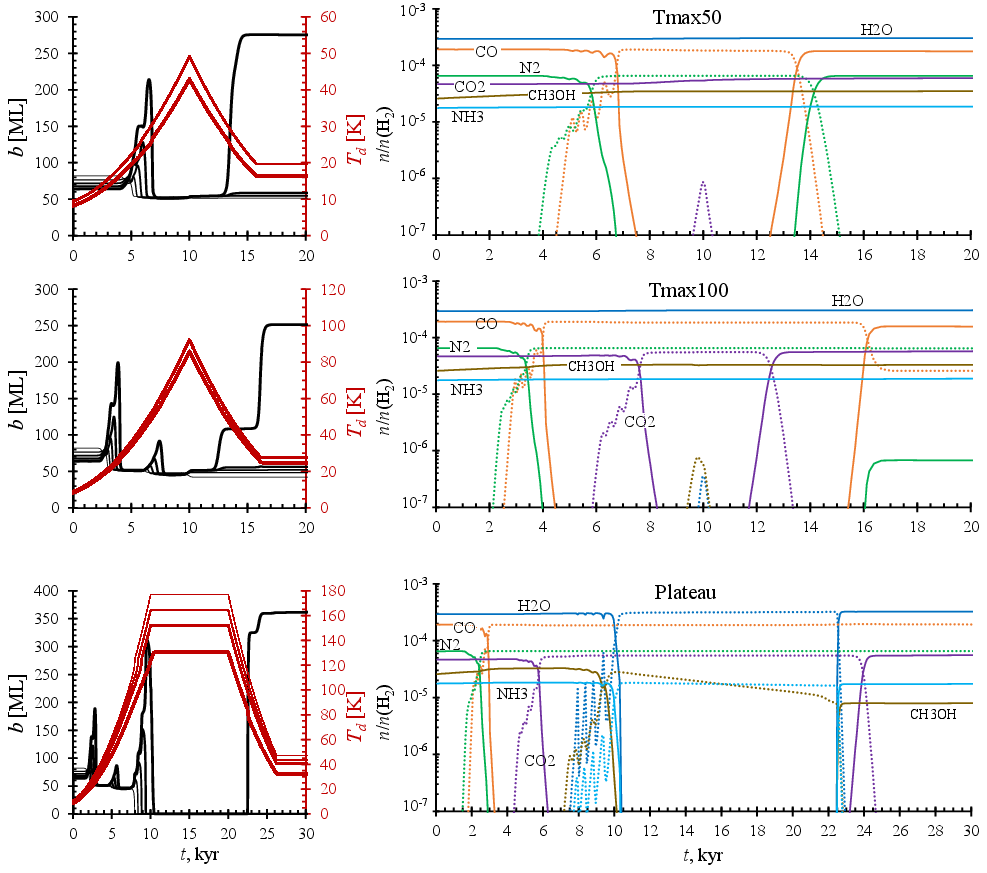}
        \vspace{-15.5cm}
        \caption{Effects of changes related to $T_{\rm max}$ on the ice sublimation and re-freeze with models {\f Tmax50} (top) and {\f Tmax100} (middle) with $T_{\rm max}$ 50 and 100\,K, respectively. The bottom plots show the results from Model {\f Plateau} with $t_{\rm plat}=10$\,kyr. Left: Ice thickness (black lines) and grain temperature (red; the thicker lines show larger grains). Right: $n/n\rm(H_2)$ of major species in gas (dotted) and ice (solid lines). This is to be compared with Model {\f Standard} in Fig.~\ref{fig-dens}, where $T_{\rm max}=150$\,K and $t_{\rm plat}=0$. The scale changes for the secondary $y$-axis ($T_d$) in the left plots.}
        \label{fig-Tmax}
\end{figure*}

Different trajectories from the envelope to the disk can have different $T_{\rm max}$. Our default value of 150\,K is among the highest \citep{Drozdovskaya14,Furuya17} and was chosen because water ice is sublimated from most grain populations at this temperature, which allows the transfer of most of the ice mass between the grain-size bins. Lower maximum gas temperatures of 50 and 100\,K were considered in models {\f Tmax50} and {\f Tmax100}, respectively, at the same $t_{\rm heat}$ and $t_{\rm cool}$.

Figure~\ref{fig-Tmax} shows that only the CO and N$_2$ ices sublimate in Model {\f Tmax50}. Both molecules also re-freeze onto the coldest grain-size bin surface before $T_{\rm end}=20$\,K is reached. The ice thickness for the 0.232\,$\rm\mu$m grains changes from 64\,MLs initially to 276\,MLs at the end of the simulation. In Model {\f Tmax100}, CO$_2$ ice also sublimates and re-accumulates, while N$_2$ ice does not re-freeze at $T_{\rm end}=30$\,K, and finally, $b_{0.232}=251$\,MLs.

The {\f Tmax50} and {\f Tmax100} models differ in another way. The heating rate is lower by factors of 3 and 2, respectively, relative to the {\f Standard} model. This is because $T_{\rm max}$ is lower, while $t_{\rm heat}$ remains constant. Table~\ref{tab-50} shows that this change has no substantial effect on $T_{\rm des}$ for the N$_2$ and CO ices, while for CO$_2$ ice, $T_{\rm des}$ is lower by 5\,K.

A third model that considered changes in the heating timescales is {\f Plateau}, where we assumed that the parcel of infalling matter spends time $t_{\rm plat}=10$\,kyr at $T_{\rm max}=150$\,K. This simple change allowed us to investigate the redistribution of ices when all major volatiles were sublimated from all grain-size bins. This scenario never occurs in Model {\f Standard}. The first 10\,kyr of {\f Plateau} are a copy of the same period in Model {\f Standard}. However, when the $T_{\rm max}$ plateau starts, H$_2$O ice is depleted below the 1\,ML levels already during the first 500 years. The only other major change that occurs during $t_{\rm plat}$ is the destruction of CH$_3$OH in the gas \citep[e.g.,][]{Millar91}. Upon re-freeze, the abundance of CH$_3$OH ice is restored to only 31\,\% of its initial abundance (i.e. $T_{\rm re,CH3OH}$ is never reached). This decline of methanol is faster than that in the model of \citet{Garrod13}, for example, because the CR ionization rate $\zeta_{\rm CR}$ in our models is higher and because we used a different gas-phase chemical network.

The re-accumulation of the H$_2$O, NH$_3$, and CO$_2$ ices is more complete than that of CH$_3$OH in Model {\f Plateau}, and similarly to the case of {\f Standard}, it exclusively occurs on the large, cold grains, where it reaches final abundances similar to those of Model {\f Standard}. The final $b_{\rm 0.232}$ is 362\,MLs. Figure~\ref{fig-Tmax} shows that ice re-acumulation occurs about 2250\,yr after the end of the $T_{\rm max}$ plateau stage, when $T_d$ for the large grains is only 84\,K (corresponding to the re-accumulation gas temperature $T_{\rm re,H2O}=96$\,K). This $T_d$ is significantly below the sublimation temperature of H$_2$O ice ($T_d\approx100$\,K). The ice accumulation occurs late because hydrogen bonds between ice species and bare-grain surfaces are lacking, similarly to models {\f Dens6} and {\f Dens7} (Sect.~\ref{r-dens}).

\subsection{Irradiation}
\label{r-irr}

\begin{figure*}
\includegraphics{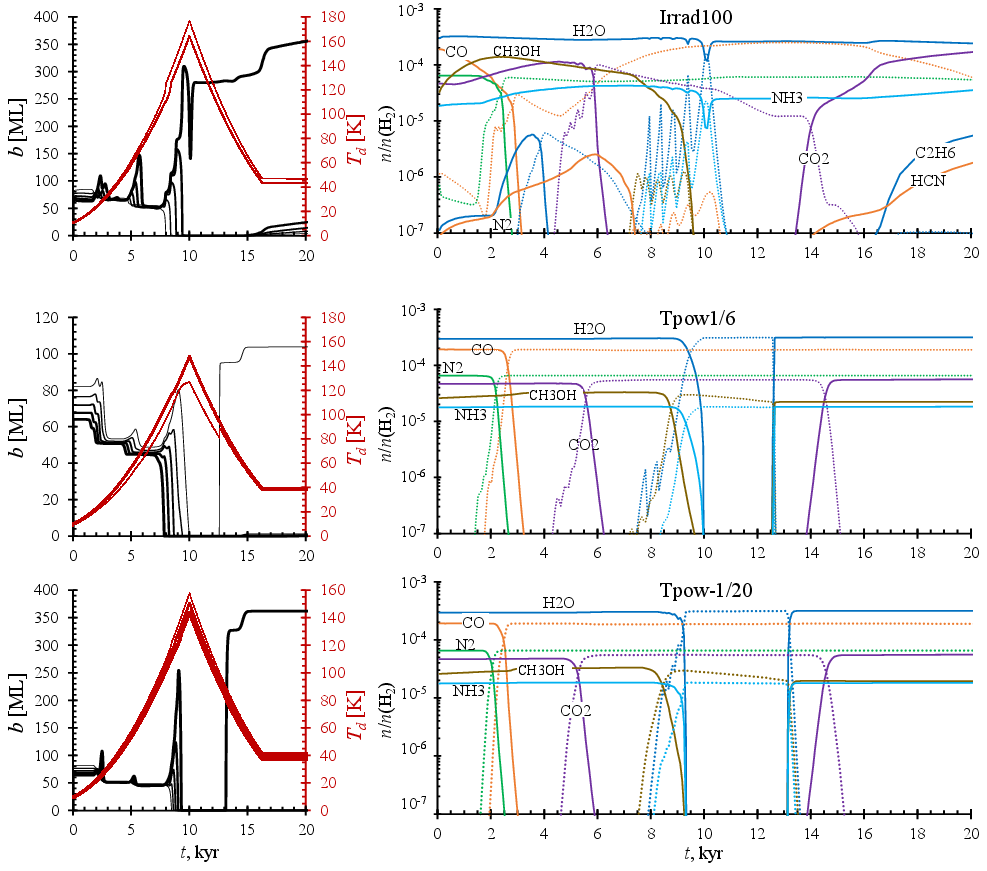}
        \vspace{-15.5cm}
        \caption{Effects of a 100-fold increase in irradiation with Model {\f Irrad100} (top) and the reverse grain temperature trend with Model {\f Tpow1/6} (bottom) on ice loss and re-accumulation. Left: Ice thickness (black lines) changes with $T_d$ (red). The thicker lines show larger grains. Right: Ice (solid lines) and gas-phase (dotted lines) abundances of the major ice species relative to H$_2$.} This is to be compared with Model {\f Standard} in Fig.~\ref{fig-dens}.
        \label{fig-citi}
\end{figure*}

With $I$ we regulated the intensity of ISRF and CRs. The value of $I$ was assumed to be unity for the {\f Standard} model, where $A_V=10$\,mag and $\zeta_{\rm CR}=4.1\times10^{17}$\,s$^{-1}$ \citep[\citetalias{K24}]{Ivlev15}. A detailed investigation of changes in radiation for infalling envelope material \citep[see e.g.][]{Drozdovskaya14} is beyond the scope of this study. Our aim here was to determine the overall trend for the ice evolution under elevated irradiation. Therefore, highly irradiated conditions were represented with a single model that differed from {\f Standard} with $I$ set to 100. This value shows significant chemical effects in the solid and gas phases, but it is not excessively destructive for the 10$^3$..10$^4$\,yr long chemical timescales considered in the simulations. In the case of ISRF, $I=100$ was achieved by lowering $A_V$ from 10 to 5, while a 100-fold increase in CRs is expected to mimic some effects of energetic particles from the protostar. The same increase was also attributed to CR-induced photons.

Figure~\ref{fig-citi} and Table~\ref{tab-50} show that evolution of the amount of ice and its thickness at the end of the simulation in Model {\f Irrad100} is generally similar, although the underlying chemical processes differ significantly. The abundance of the ice molecules is reduced by three processes: thermal desorption, photodesorption, and ice photoprocessing. The latter results in fragments of photodissociated solid-phase species that combine into new molecules, part of which is desorbed. Thus, the loss of ices in this model can no longer be described as simply `sublimation'. Under the strong irradiation, the chemistry of major elements and species changes, as summarised below.
\begin{itemize}
\item Carbon and oxygen. Initially, CO ice is rapidly destroyed by photoprocesses and is already depleted at 18\,K, when CO sublimation does not yet occur. H$_2$O ice and CO gas are combined into CO$_2$ and CH$_3$OH ices, which both sublimate and are mostly destroyed in the gas. During the cooling stage, only CO$_2$ ice is restored, and at the end time of 20\,kyr, the abundance of CO$_2$ has become comparable to that of H$_2$O ice, with $n/n({\rm H_2})$ of $1.7\times10^{-4}$ and $2.4\times10^{-4}$, respectively.
\item Organic molecules. The synthesis of CH$_3$OH via hydrogenation of CO becomes ineffective when $T_d>24$\,K because atomic H, produced by H$_2$O photodissociation or adsorbed from the gas-phase, sublimates too rapidly. CH$_3$OH is fully desorbed and then destroyed by the gas-phase chemistry, and does not re-form onto grains during the second half of the simulation. The gas-phase chemistry produces hydrocarbons, such as ethane C$_2$H$_6$, and those with desublimation temperatures above 40\,K accumulate onto the grain surfaces.
\item Nitrogen. As N$_2$ sublimates, N is slowly accumulated into the much less volatile NH$_3$ ice \citep{KK19}. During the cooling stage and especially towards the end of the simulation, other ice species, such as HCN and HC$_3$N, become abundant, although most N atoms still go into NH$_3$ ice.
\end{itemize}

During the simulation end phase with constant $T_{\rm end}=40$\,K, the ice thickness continues to rise steadily and reaches $b_{0.232}=356$\,MLs for the large grains and $<25$\,MLs for other grain-size bins. This continued grain growth occurs because the major gas-phase species CO and N$_2$ are constantly directly destroyed by radiation or ions, such as He$^+$, which is produced by CRs. The ionization induced by ISRF is less significant because of molecular (self-) shielding effects. Some of the resulting atoms and atomic ions recombine into less volatile molecules that stick to the grain surfaces, where they continue to increase the ice thickness.

\subsection{Grain temperature trend}
\label{r-Tpow}

We adopted a default value of $y=-1/6$ in the {\f Standard} model, that is, small grains with a large total surface area were hotter than large grains. This trend is supported by \citet{Draine03,Cuppen06}, and \citet{Krugel07}. The opposite trend \citep[e.g.,][]{Heese17} with $y=1/6$ was considered with Model {\f Tpow1/6}. \citetalias{KF24} concluded that the least likely scenario is that all grains have similar temperatures because $T_d$ is affected not only by the grain size, but even more so by their material \citep[see Sect.~\ref{intrd} and][]{Woitke24}.

Figure~\ref{fig-citi} shows that the main difference in the results of Models {\f Standard} and {\f Tpow1/6} is the early and rapid ice sublimation, followed by a late and rapid re-accumulation in {\f Tpow1/6}. This is shown quantitatively with the sublimation temperatures in Table~\ref{tab-50}. All ice is sublimated at 150\,K. The first ice layer reappears on the small grains when their $T_d$ drops to 81\,K (at $T_{\rm gas}=95$\,K). Similarly to Models {\f Dens6}, {\f Dens7}, and {\f Plateau}, the ice re-accumulation is delayed by the assumed non-existence of hydrogen bonds on the bare-grain surface. After the first ice ML has formed, the small grain-size bin adsorbs all the hydrogen-bonding less volatile ices within 10\,yr.

The accumulation of all the hydrogen-bonded species increases the radius of the $a=0.037\,\rm\mu$m grains to $a+b=0.067\,\rm\mu$m, which is higher than the second-smallest grain-size bin with $a=0.058\,\rm\mu$m. With $y>0$ and size growth, the $T_d$ of the small grains suddenly rises from 81 to 88\,K before it continues to decline, as programmed in the code (Sect.~\ref{grd}). The rise delays the accumulation of the next volatile, CO$_2$, which explains why $T_{\rm re,CO2}=63$\,K in Model {\f Tpow1/6} is lower than that for Model {\f Standard}, where it is 72\,K. (We recall that $T_{\rm des}$ and $T_{\rm re}$ are gas temperatures that are indicative of the assortment for $T_d$ of the five grain-size bins.) Similarly to Model {\f Standard}, the CO and N$_2$ ices do not re-accumulate in Model {\f Tpow1/6}.

While the temperature calculations for grains with different sizes and compositions indicate a significant spread in $T_d$ (see Sect.~\ref{micr}), we also explored Model {\f Tpow-1/20} with a narrow $T_d$ range with $y$ taken to be -1/20, similarly to Models {\f Dens10} and {\f Dens12}. These similar grain temperatures may occur when all grains happen to share a similar composition and size. A narrow $T_d$ spread reduces the possibilities for a redistribution of the ice mass between the grain-size bins (Sect.~\ref{r-dens}). Figure~\ref{fig-citi} shows that a full ice removal occurs in Model {\f Tpow-1/20}. The $T_{\rm des}$ and $T_{\rm re}$ values are lower than those of Model {\f Standard} because of the higher minimum $T_d$ for large grains. During the cooling stage, the recovery of H$_2$O, NH$_3$, and CH$_3$OH ices occurs, and the additional delay is induced by the bare-grain surface, which does not participate in hydrogen bonds and thus has lower thermal desorption temperatures for these species (Sect.~\ref{r-dens}).

\subsection{Distribution of the ices on the grains}
\label{r-ini}

\begin{figure*}
\includegraphics{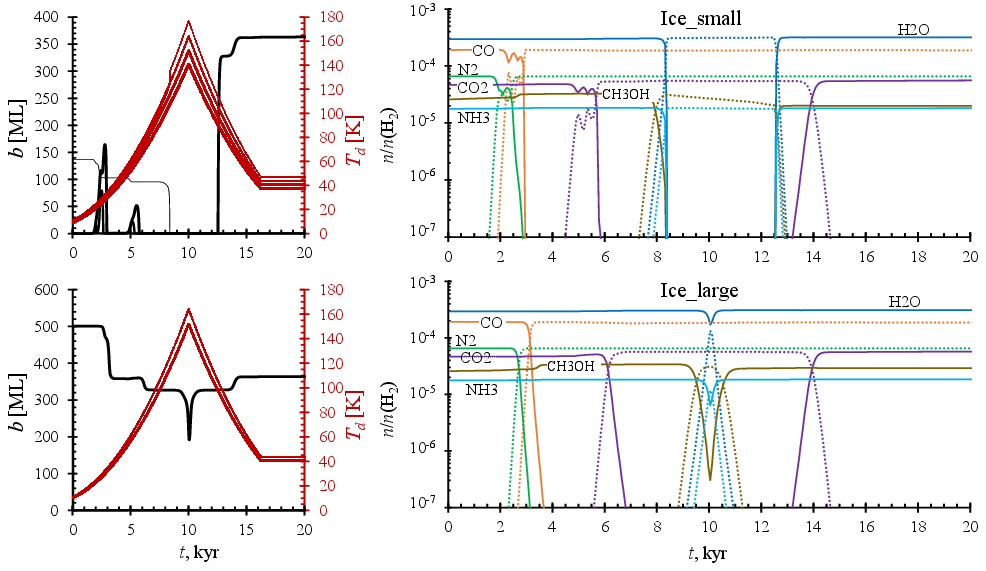}
        \vspace{-20.5cm}
        \caption{Results of Model {\f Ice\_small}, where all ice is on the surface of $a=0.037\,\rm\mu$m grains (top) and Model {\f Ice\_large} with ice on the 0.232\,$\rm\mu$m grains (bottom). Left: Ice thickness (black lines) changes with $T_d$ (red). The thicker lines show larger grains. Right: Ice (solid lines) and gas-phase (dotted lines) abundances of the major ice species relative to H$_2$. This is to be compared with Model {\f Standard} in Fig.~\ref{fig-dens}.}
        \label{fig-ini}
\end{figure*}

Interstellar ice may be unevenly distributed among grains, and some grain-size bins may have thicker ice at the expense of others \citep{Iqbal18,Silsbee21}. In order to determine whether this phenomenon affects the distribution of ice among the grain-size bins after a heating-cooling cycle, we considered two models in which all ice was either located on the surface of the smallest 0.037\,$\rm\mu$m grains (Model {\f Ice\_small}) or of the largest 0.232\,$\rm\mu$m grains (Model {\f Ice\_large}). In terms of the ice distribution, these two models can be viewed as fictional illustrating cases.

Figure~\ref{fig-ini} shows that the extreme initial distribution of ices on the surface of a single grain-size bin in Models {\f Ice\_small} and {\f Ice\_large} does not change the end result, the ice accumulation onto the coldest grains, that we obtained with the other models. The phase transition in Model {\f Ice\_large} is similar to that in a single-grain model. All ices start, sublimate, and re-adsorb onto the surfaces of the coldest grain-size bin. No appreciable ice layer ever appears on the surface of other grain-size bins, where the ice thickness does not exceed 0.4\,MLs at any point in time. The overall evolution of the chemical abundances closely resembles that of Model {\f Standard}, except for the lack of the seesaw pattern in the gas phase (Sect.~\ref{r-std}). Table~\ref{tab-50} shows that the $T_{\rm des}$ values are higher by a few degrees than those of the {\f Standard} model because no previous sublimation from warmer grains was possible in Model {\f Ice\_large}. After $T_{\rm max}$, all information on previous ice distribution is lost, and Models {\f Standard} and {\f Ice\_large} show very similar results in the cooling stage.

The initial distribution of ices in Model {\f Ice\_large} resembles the final distribution in Model {\f Standard}. Moreover, the final distribution of the ices of Model {\f Ice\_large} is similar to the initial distribution of this same model. In other words, Model {\f Ice\_large} can be viewed as the next heating-cooling cycle of a gas parcel after the cycle of Model {\f Standard}. This second cycle again results in ice that is exclusively accumulated on the surface of the coldest grain-size bin. This statement excludes the hyper-volatile species that do not accumulate because $T_{\rm end}=40$\,K prevents their adsorption. Nevertheless, Model {\f Tmax50} indicates that these species re-accumulate in a similar manner, but onto the coldest grains.

Model {\f Ice\_small} displays a more complex evolution than {\f Ice\_large}, which nevertheless arrives at the same end result. The hyper-volatile species and CO$_2$ sublimate from the small 0.037\,$\rm\mu$m and re-freeze onto the surface of the larger, colder grains before they sublimate fully. This is similar to Model {\f Standard}, where a major part of the ice also is on the 0.037\,$\rm\mu$m grains. This process is not repeated for H$_2$O, NH$_3$, and CH$_3$OH because the larger and colder grains have bare, carbonaceous surfaces by the definition of Model {\f Ice\_small}. This means that $E_D$ for these species is significantly lower, and they cannot accumulate. Therefore, full ice sublimation occurs in Model {\f Ice\_small} already when H$_2$O and the associated ices disappear from the 0.037\,$\rm\mu$m grains at $T_{\rm gas}=114$\,K and $T_{0.037}=134$\,K. This results in a delayed re-accumulation of the ices, as discussed in Sect.~\ref{r-dens}. The early desorption and late adsorption, compared to Model {\f Standard}, can be seen quantitatively with the $T_{\rm des}$ and $T_{\rm re}$ values in Table~\ref{tab-50}.

\subsection{Summary}
\label{summm}

The results of Model {\f Standard} and of most of the other models show that, first, during the heating period, the molecules that were sublimated from warmer grains re-settle into the coldest grain-size bin before full sublimation occurs. Second, during the cooling period, ice irreversibly and exclusively re-accumulates into the coldest bin. This redistribution of ice into the coldest grain-size bin is valid for gas densities from $10^7$\,cm$^{-3}$, which is characteristic of dense, heated envelopes to $10^{12}$\,cm$^{-3}$ in PPD interior. The large 0.232\,$\rm\mu$m grains have 12\,\% of the entire bare-grain surface area in the model. For the redistribution to occur, the majority of the heated grains must exceed the sublimation temperature of the ice species. The hyper-volatile ices N$_2$ and CO disappear when $T_d$ exceeds $\approx$34\,K, and they are likely redistributed for all infall trajectories. For CO$_2$ ice, this temperature is $>60$\,K, while for less volatile ices, such as H$_2$O, it is $\approx$124\,K. These $T_d$ correspond to the temperature of coldest available grains, which means that $T_{\rm gas}$ will always be higher.

Under conditions of high irradiation, CO and H$_2$O ices are converted into CO$_2$. Therefore, more of the ice mass can sublimate and re-freeze at medium $T_d$ of $\approx$60\,K (up to 73\,K in Model {\f Dens12}). Irradiation does not change the general pattern of ice redistribution onto the coldest grains. This is also true for models with altered grain-size $T_d$ distributions, for instance, for Model {\f Tpow1/6}, where the largest grains have the highest $T_d$ and Model {\f Tpow-1/20} with a narrow $T_d$ distribution. Unlike Model {\f Standard}, full ice sublimation occurs in both models, and the redistribution occurs when the infalling parcel of matter cools again.

A final aspect is that the re-freeze of the H$_2$O-like ices occurs at much lower $T_d$ in cases of full ice sublimation. This is because the bare-grain surface was assumed to be carbonaceous and does not form hydrogen bonds in our model. Therefore, $T_{\rm re}$, which is necessary for re-forming the first full ice ML, is lower by about 30...40\,K than $T_{\rm des}$. In a real protostellar envelope, the surfaces of the refractory grains can also consist of silicates or other materials. Along with the $T_d$ differences between grain materials \citepalias{KF24}, this means that the grain surface material is another variable that can promote accumulation of ice onto a minority of grains with specific properties.

\section{Conclusions}
\label{cncl}

To summarise the results, the selective accumulation of ice onto the surface of the coldest-grain population during a heating-cooling cycle is a strong rule. It cannot be easily changed by variations in the physical or chemical conditions of matter falling onto a PPD from a protostellar envelope. This means that most grains enter the PPD ice poor, while a small minority are ice rich. This uneven ice distribution can affect the grain coagulation, which is more likely and at higher collision speeds, when the grains are covered with an ice layer \citep{Gundlach15,Nietiadi17,Nietiadi20}. As far as we know, the specific case of collisions between ice-covered and bare refractory grains has not yet been investigated. The relative grain velocities can also be significantly affected by their size differences \citep{Ormel09,Silsbee20}, which can be exaggerated or reduced by the selective accumulation of ice.

In prestellar cores, grains grow by accumulating icy mantles \citep{Whittet01}, while rapid further grain growth by coagulation is observed already in protostellar envelopes \citep{Miotello14,Testi14}. Coagulation is affected by the dual nature of icy and bare grains that are present in the same place and at the same time. Heating events affecting PPDs, such as the event observed by \citet{Burns23}, can also be expected to irreversibly produce dual grain populations (bare and ice rich), which affects the behaviour of grains in the disk, as described by \citet{Podolak11}. The spread-out snowlines in PPDs, predicted by \citet{Gavino21}, likely do not occur, as the ices always sublimate from and freeze out onto the same population of the coldest grains as in Model {\f Ice\_large}.

The full sublimation of ice and its subsequent re-accumulation will result in fully stratified icy mantles with species of increasing volatility towards the surface. This aspect was not simulated in this model. This ice structure facilitates the rapid sublimation of ice components with no need to diffuse through layers of less volatile ices. Moreover, the segregation into relatively pure ices may hamper the synthesis of some complex molecules, including organics. For example, one route for CH$_3$OH production is by photoprocessing a H$_2$O-CO icy mixture, which cannot be expected to be active in stratified ices.

\begin{acknowledgements}
This research is funded by the Latvian Science Council grant ``Desorption of icy molecules in the interstellar medium (DIMD)'', project No. lzp-2021/1-0076. I am grateful for the support of Ventspils City Council. This research has made use of NASA’s Astrophysics Data System. I thank the anonymous referee for efforts that significantly improved the paper.
\end{acknowledgements}

   \bibliographystyle{aa}
   \bibliography{planets}

\end{document}